\documentclass[a4paper,12pt]{article}
\newtheorem{definition}{{\bf Definition}}[section]
\newtheorem{theorem}[definition]{{\bf Theorem}}
\newtheorem{lemma}[definition]{{\bf Lemma}}

\newtheorem{corollary}[definition]{{\bf Corollary}}

\def\<{\langle}
\def\>{\rangle}
\def\im{{\rm i}}

\def\H{{\cal H}}

\def\card{{\rm card}}
\def\a{{\bf a}}
\def\rank{{\rm rank}}
\def\Ran{{\rm Ran}}
\def\e{{\bf e}}

\def\proof{\noindent{Proof. }}
\def\endproof{\hfill $\square$ \vspace{10pt}}
\usepackage{amscd,amsmath,amssymb,graphicx,cite}

\begin{document}

\title{Parameterization of quantum walks on cycles}
\author{Shuji Kuriki\footnote{Department of Mathematics, Faculty of Engineering, 
Shinshu University,
4-17-1 Wakasato, Nagano 380-8553, Japan}, \quad 
Md Sams Afif Nirjhor\footnote{Graduate School of Information Sciences, 
Tohoku University,
Aoba-ku, Sendai, 980-8579, Japan}, \quad
Hiromichi Ohno\footnotemark[1]}

\date{}

\maketitle

\begin{abstract}
This study investigate the unitary equivalence classes of quantum walks on cycles.
We show that unitary equivalence classes of quantum walks on a cycle with $N$ vertices are 
parameterized by $2N$ real parameters.
Moreover, the ranges of two of the parameters are restricted, 
and the ranges depend on the parity of $N$.\\
\end{abstract}


\section{Introduction}
Quantum walks are analogous to classical random walks. 
They have been studied in various fields, 
such as quantum information theory and quantum probability theory. 
A quantum walk is defined by a pair $(U, \{H_x\}_{x\in V})$, in
which $V$ is a countable set, $\{H_x\}_{x\in V}$ is a family of separable Hilbert spaces,
and $U$ is a unitary operator on $\H = \bigoplus_{x\in V} \H_x$ \cite{SS2}.
In this paper, we discuss
quantum walks on a cycle, in which $V = \{1,2, \ldots, N\}$ and $\H_x = {\mathbb C}^2$.
These have been the subject of some previous studies \cite{GZ, D, IK, HK}.

It is important to clarify when two quantum walks are 
unitarily equivalent in the sense of \cite{O, SS2}. 
If two quantum walks are unitarily equivalent, 
many properties of their quantum walks are the same. 
For example, digraphs, dimensions of Hilbert
spaces, spectrums of unitary operators, probability distributions of quantum
walks, etc. would be the same for each quantum walk. 
The aim of this paper is to determine the unitary equivalence classes of 
quantum walks on cycles.
Then, we only need to study representatives of unitary equivalence classes
to know the above properties.

In the previous papers \cite{O, O2, O3}, we considered unitary equivalence classes
of one-dimensional and two-dimensional quantum walks.
Unitary equivalence classes of translation-invariant
one-dimensional quantum walks were also investigated in \cite{GKD}.

In Sect. \ref{sec2}, we show a natural expression of quantum walks with some conditions.
After that, we consider unitary equivalence of such quantum walks.
The results in Sect. 2 are similar to those in \cite{O, IKB}, but improved  a little.

In Sect. 3, we prove that unitary equivalence classes of quantum walks on a cycle with
$N$ vertices are parameterized by $2N$ real numbers. 
This parameterization is similar to that of one-dimensional quantum walks,
because we need two parameters for each vertex to parameterize 
the unitary equivalence classes of one-dimensional quantum walks \cite{O2}. 
On the other hand, ranges of two of the parameters are restricted, 
and the two parameters go to zero when $N$ goes to infinity.
Moreover, the ranges depend on the parity of $N$.
These properties are not seen in the cases of one-dimensional and two-dimensional quantum walks.


\section{Natural expression of quantum walks}\label{sec2}

In this section, we present a natural expression of quantum walks with some conditions.
We also consider unitary equivalence of such quantum walks.

Let $V$ be a countable set.
For each $x \in V$,  $\H_x = {\mathbb C}^{k_x}$ is a finite dimensional Hilbert space,
and $P_x$ is a projection from $\H = \bigoplus_{y\in V} \H_y$ onto $\H_x$.
A unitary $U$ on $\H$ is called a quantum walk \cite{SS2, O}.
Given a quantum walk $U$ on $\H$, we can construct a multidigraph $G_U = (V, D_U)$.
For vertices $x, y \in V$, the number of directed edges from $y$ to $x$ is denoted by
$\card (x,y)$; i.e.,
\[
\card (x,y) = \card \{ \a \in D_U \colon t(\a ) =x , o(\a) =y \},
\]
where $o(\a)$ and $t(\a)$ are the origin and terminus of the directed edge $\a$, respectively,
and card indicates the cardinal number of a set.
We define the number of directed edges from $y$ to $x$ by
\[
\card (x,y) = \rank P_x U P_y.
\]
Then, a multidigraph $G_U = (V, D_U)$ is  called a multidigraph of the quantum walk $U$.
We will write $G=G_U$ and $D=D_U$, when there is no confusion.

We consider quantum walks which satisfy one of the following conditions: for all $x \in V$,
\begin{eqnarray}
\card \{\a \in D \colon o(\a) =x \} = \dim \H_x, \label{cond:origin}\\
\card \{\a \in D \colon t(\a) = x \} = \dim \H_x. \label{cond:terminus}
\end{eqnarray}
We prepare two lemmas.

\begin{lemma}\label{lemma2.1}
If a quantum walk $U$ satisfies \eqref{cond:origin}, then for each $x \in V$,
$\Ran (UP_x) = \bigoplus_{y \in V } \Ran (P_y U P_x)$.
Moreover, $U|_{\H_x}$ is a unitary from $\H_x$ onto $ \bigoplus_{y \in V } \Ran (P_y U P_x)$.
\end{lemma}
\proof
For any $UP_x \psi \in \Ran (UP_x)$,
\[
UP_x \psi = \sum_{y \in V} P_y U P_x \psi \in \bigoplus_{y\in V} \Ran (P_y U P_x).
\]
Therefore, $\Ran (UP_x) \subset \bigoplus_{y \in V } \Ran (P_y U P_x)$.
This implies $\rank UP_x \le \sum_{y\in V} \rank P_y U P_x$.

By definitions and \eqref{cond:origin},
\begin{align*}
\dim \H_x &= \dim U \H_x = \rank UP_x \le \sum_{y \in V} \rank P_y U P_x 
= \sum_{y \in V} \card (y,x)\\
&= \card \{ \a \in D \colon o(\a) = x\} = \dim \H_x.
\end{align*}
Hence, $\rank UP_x = \sum_{y \in V} \rank P_y U P_x$ and therefore, 
\[
\dim \Ran (UP_x) = \sum_{y \in V } \dim \Ran (P_y U P_x) 
= \dim \bigoplus_{y \in V }  \Ran (P_y U P_x).
\]
Since $\Ran (UP_x) \subset \bigoplus_{y \in V } \Ran (P_y U P_x)$,
we obtain $\Ran (UP_x) = \bigoplus_{y \in V } \Ran (P_y U P_x)$.

By the equation $U\H_x= \Ran (UP_x) = \bigoplus_{y \in V } \Ran (P_y U P_x)$,
$U|_{\H_x}$ is a unitary from $\H_x$ onto $\bigoplus_{y \in V } \Ran (P_y U P_x)$
\endproof

\begin{lemma}\label{lemma2.2}
If a quantum walk $U$ satisfies \eqref{cond:terminus}, then $U^*$ satisfies \eqref{cond:origin}.
\end{lemma}

\proof
By the equation $\rank P_x U P_y = \rank P_y U^* P_x$ and \eqref{cond:terminus},
\begin{align*}
\dim \H_x &= \card \{\a \in D_U \colon t(\a) =x \} = \sum_{y \in V} \card (x,y)
= \sum_{y \in V} \rank P_x U P_y \\
&= \sum_{y \in V} \rank P_y U^* P_x 
= \card \{ \a \in D_{U^*} \colon o(\a)=x\}.
\end{align*}
Hence, 
$U^*$ satisfies \eqref{cond:origin}.
\endproof

The next theorem shows a natural expression of a quantum walk with the condition 
\eqref{cond:origin} or \eqref{cond:terminus}.
This result is similar to that in \cite{O, IKB}, but improved a little.

\begin{theorem}\label{thm2.3}
If a quantum walk $U$ satisfies \eqref{cond:origin} or \eqref{cond:terminus},
there exist orthonormal bases $\{\xi_\a \}_{\a \in D}$ and $\{\zeta_\a \}_{\a \in D}$
of the Hilbert space $\H$ with $\xi_\a \in \H_{t(\a)}$ and $\zeta_\a \in \H_{o(\a)}$
such that
\[
U = \sum_{\a \in D} |\xi_\a \> \< \zeta_\a|.
\]
\end{theorem}

\proof
First, we assume that $U$ satisfies \eqref{cond:origin}. 
Since $\dim \Ran(P_y U P_x)= \rank P_y U P_x = \card (y,x)$ for all $x,y \in V$, 
an orthonormal basis of $\Ran(P_y U P_x)$ is indexed by directed edges
$\{ \a \colon \a \in D, t(\a) = y, o(\a) =x \}$.
Hence, 
there is an orthonormal basis 
$\{ \xi_{\a} \colon \a \in D, t(\a) = y, o(\a) =x \}$ of $\Ran (P_y U P_x)$.
Note that $\xi_\a \in \H_y = \H_{t(\a)}$.
Then, the union
\[
\bigcup_{y \in V} \{\xi_{\a} \colon \a \in D, t(\a) = y, o(\a) =x \}
 =\{ \xi_{\a} \colon \a \in D, o(\a)= x\}
\]
is an orthonormal basis of $\bigoplus_{y\in V} \Ran (P_y U P_x)$.
Define $\zeta_\a = U^* \xi_\a$.
By Lemma \ref{lemma2.1}, $\{ \zeta_\a \colon  \a \in D, o(\a)= x\}$ is an orthonormal basis
of $\H_x = \H_{o(\a)}$. 
Then, the union
\[
\bigcup_{x\in V} \{ \zeta_\a \colon  \a \in D, o(\a)= x\} 
= \{ \zeta_\a \colon \a \in D \}
\]
is an orthonormal basis of $\H$.
Since $U$ is unitary and $\xi_\a = U\zeta_\a$, $\{\xi_\a \colon \a \in D\}$ is also
an orthonormal basis of $\H$.
Consequently, we have orthonormal bases 
$\{\xi_\a \}_{\a \in D}$ and $\{\zeta_\a \}_{\a \in D}$
of the Hilbert space $\H$ with $\xi_\a \in \H_{t(\a)}$ and $\zeta_\a \in \H_{o(\a)}$
such that
\[
U = \sum_{\a \in D} |\xi_\a \> \< \zeta_\a|. 
\]

Next, we assume that $U$ satisfies \eqref{cond:terminus}. 
By Lemma \ref{lemma2.2},
$U^*$ satisfies \eqref{cond:origin}.
Therefore,
there exist orthonormal bases
$\{\zeta_{\a} \}_{\a \in D_{U^*}}$ and $\{\xi_{\a} \}_{\a \in D_{U^*}}$
of the Hilbert space $\H$ with $\zeta_{\a} \in \H_{t(\a)}$ and $\xi_\a \in \H_{o(\a)}$
such that
\[
U^* = \sum_{\a \in D_{U^*}} |\zeta_\a \> \< \xi_\a|.
\]
The equation ${\rm rank} P_x U P_y = {\rm rank} P_y U^* P_x$ implies 
$D_{U^*} = \{ \bar\a \colon \a \in D_U\}$, where $\bar\a$ is the inverse edge of $\a$.
This allows us to change the index set from $D_{U^*}$ to $D_U$, that is,
\[
U^* = \sum_{\a \in D_{U}} |\zeta_{\a} \> \< \xi_{\a}|
\]
with $\xi_{\a} \in \H_{t(\a)}$ and $\zeta_\a \in \H_{o(\a)}$.
Consequently, we have
\[
U = \sum_{\a \in D_{U}} |\xi_\a \> \< \zeta_\a|,
\]
where $\{\xi_{\a} \}_{\a \in D_{U}}$ and $\{\zeta_{\a} \}_{\a \in D_{U}}$
are orthonormal bases of $\H$ with 
$\xi_{\a} \in \H_{t(\a)}$ and $\zeta_\a \in \H_{o(\a)}$.
\endproof

As a corollary of this theorem, we have the following.

\begin{corollary}
For a quantum walk $U$, the conditions  \eqref{cond:origin} and \eqref{cond:terminus}
are equivalent.
\end{corollary}

\proof
Assume that $U$ satisfies \eqref{cond:origin}. By Theorem \ref{thm2.3},
there exist orthonormal bases $\{\xi_\a \}_{\a \in D}$ and $\{\zeta_\a \}_{\a \in D}$
of the Hilbert space $\H$ with $\xi_\a \in \H_{t(\a)}$ and $\zeta_\a \in \H_{o(\a)}$
such that
\[
U = \sum_{\a \in D} |\xi_\a \> \< \zeta_\a|.
\]
Since $\{\xi_\a\}_{\a \in D}$ is an orthonormal basis of $\H$ and $\xi_\a$ is in $\H_{t(\a)}$,
for each $x \in V$, the set
\[
\{ \xi_\a  \colon \a \in D, t(\a) = x\}
\]
is an orthonormal basis $\H_x$. Therefore, $U$ satisfies \eqref{cond:terminus}.

On the other hand, assume that $U$ satisfies \eqref{cond:terminus}.
By Lemma \ref{lemma2.2}, $U^*$ satisfies \eqref{cond:origin} and therefore, 
$U^*$ satisfies \eqref{cond:terminus}.
Again, by Lemma \ref{lemma2.2}, $U$ satisfies \eqref{cond:origin}.
\endproof

Now, we consider unitary equivalence of quantum walks.
We recall the definition of unitary equivalence of quantum walks.

\begin{definition}
Quantum walks $U_1$ and $U_2$ on $\H = \bigoplus_{x \in V} \H_x$ are unitarily equivalent
if there exists a unitary $W =\bigoplus_{x\in V} W_x$ on $\H$ such that
\[
WU_1 W^* = U_2.
\]
\end{definition}

In a natural expression in Theorem \ref{thm2.3}, we need two
orthonormal bases $\{ \xi_\a \}$ and $\{\zeta_\a\}$.
Considering unitary equivalence of quantum walks
with the conditions \eqref{cond:origin} and \eqref{cond:terminus},
we can disappear one of them.
$\{ \e_i^x \}_{i=1}^{k_x}$ denotes a canonical basis of $\H_x = {\mathbb C}^{k_x}$.

\begin{theorem}\label{thm2.6}
If $U$ satisfies \eqref{cond:origin} or \eqref{cond:terminus},
there exists an orthonormal basis $\{\zeta_{i}^x  \colon i=1, \ldots, k_x, x\in V\} $ 
of $\H$
such that
$U$ is unitarily equivalent to
\[
U_{\zeta} = \sum_{x \in V} \sum_{i = 1}^{k_x} |\e_{i}^x\> \<\zeta_{i}^x|,
\]
and $\zeta_i^x$ is in $\H_y$ for some $y$ which satisfies $(x,y) \in D$.
\end{theorem}

\proof
By Theorem \ref{thm2.3}, 
there exist orthonormal bases $\{\xi_\a \}_{\a \in D}$ and $\{\zeta_\a \}_{\a \in D}$
of $\H$ with $\xi_\a \in \H_{t(\a)}$ and $\zeta_\a \in \H_{o(\a)}$
such that
\[
U = \sum_{\a \in D} |\xi_\a \> \< \zeta_\a|.
\]
Since $\card \{\a \in D \colon t(\a) = x\} = \dim \H_x = k_x$, we can write
\[
\{\xi_\a \colon \a \in D, t(\a) =x \} = \{ \xi_i^x\}_{i=1}^{k_x} \quad {\rm and} \quad
\{\zeta_\a \colon \a \in D, t(\a) =x \} = \{ \eta_i^x\}_{i=1}^{k_x}.
\]
Note that $\{ \xi_i^x\}_{i=1}^{k_x}$ is an orthonormal basis of $\H_x$.
Then, $U$ can be written as
\[
U= \sum_{x\in V} \sum_{i=1}^{k_x} |\xi_i^x \>\<\eta_i^x|.
\]
Define a unitary $W$ by
\begin{equation}\label{eq2.6}
W = \bigoplus_{x \in V} \sum_{i=1}^{k_x} |\e_{i}^{x}\> \< \xi_i^x |. 
\end{equation}
Then,
\[
WUW^* =  \sum_{x\in V} \sum_{i=1}^{k_x} |W \xi_i^x \> \< W \eta_i^x|
=\sum_{x\in V} \sum_{i=1}^{k_x} |\e_i^x \> \< \zeta_i^x|,
\]
where $\zeta_i^x = W \eta_i^x$.
By definition, $\zeta_i^x$ is in $\H_y$ for some $y$ which satisfies $(x,y) \in D$.
\endproof


\section{Unitary equivalence classes of quantum walks on cycles}\label{sec3}

In this section, we consider unitary equivalence classes of quantum walks on cycles.
The vertex set is $V = \{ 1, 2, \ldots, N \}$ $(N \ge 3)$.
For each $x \in V$, $\H_x = {\mathbb C}^2$.
We define $\H = \bigoplus_{x=1}^N \H_x$.
$P_x$ is a projection from $\H$ onto $\H_x$,
and $\{ \e_{1}^x, \e_2^x\}$ is a canonical basis of $\H_x = {\mathbb C}^2$.

\begin{definition}
A unitary $U$ on $\H$ is called a quantum walk on a cycle if
\[
\rank P_{y} U P_x = 
\begin{cases}
1 \quad y = x\pm 1 \\
0 \quad {\rm otherwise}
\end{cases}
\]
for all $x \in V$, where we define $N+1 =1$ and $0 = N$ with respect to $x \pm 1$.
\end{definition}

By definition, a quantum walk $U$ on a cycle satisfies \eqref{cond:origin}.
Hence, by Theorem \ref{thm2.6}, there exists 
an orthonormal basis $\{\zeta_1^x, \zeta_2^x \}_{x\in V}$ of $\H$ such that
$U$ is unitarily equivalent to
\[
U_\zeta = \sum_{x\in V} \left( |\e_1^x\> \<\zeta_1^x| + |\e_2^x\>\< \zeta_2^x | \right).
\]
Here, $\zeta_i^x$ is in $\H_y$ for some $y$ which satisfies $(x,y) \in D$.
For any $x \in V$, $(x, y)$ is in $D$ if and only if $y = x\pm 1$ by definition.
Therefore, $\zeta_i^x$ is in $\H_{x+1}$ or $\H_{x-1}$.
We assume that $\zeta_1^x \in \H_{x-1}$ and $\zeta_2^x \in \H_{x+1}$
without loss of generality.
We set $\zeta_1^{x+1} = \eta_1^x$ and $\zeta_2^{x-1} = \eta_2^x$.
Then, $\{ \eta_1^x , \eta_2^x\}$ is an orthonormal basis of $\H_x$, and
\begin{equation}\label{eq3.1}
U_\zeta = \sum_{x\in V} \left( |\e_1^x\> \<\eta_1^{x-1}| + |\e_2^x\>\< \eta_2^{x+1} | \right)
= \sum_{x\in V} \left( |\e_1^{x+1}\> \<\eta_1^{x}| + |\e_2^{x-1}\>\< \eta_2^{x} | \right).
\end{equation}
Since $\{ \eta_1^x , \eta_2^x\}$ is an orthonormal basis of $\H_x$,
we can write
\[
\eta_1^x = r_x e^{\im a_x} \e_1^x + \sqrt{1-r_x^2} e^{\im b_x} \e_2^x, \quad 
\eta_2^x = \sqrt{1-r_x^2} e^{\im c_x} \e_1^x + r_x e^{\im d_x} \e_2^x
\]
for some $0 \le r_x \le 1$ and $a_x, b_x, c_x, d_x \in {\mathbb R}$ with
$a_x -c_x = b_x-d_x +\pi$ $({\rm mod} \ 2\pi)$, where $\im = \sqrt{-1}$.
We will use $s_x = \sqrt{1- r_x^2}$ for short, and omit $({\rm mod} \ 2\pi)$ if there is no confusion.

We prepare three lemmas to get the unitary equivalence classes of quantum walks on cycles.

\begin{lemma}\label{lemma3.2}
The set $\frac{4\pi }{N} {\mathbb Z}$ is equal to the following set in modulo $2\pi$:
\begin{align*}
&\left\{ \frac{4\pi m}{N} \colon 0 \le m \le \frac{N}{2}-1 \right\} 
\quad ({\rm when}\ N \ {\rm is \ even}), \\
&\left\{ \frac{2\pi m}{N} \colon 0 \le m \le N-1 \right\}  \quad ({\rm when}\ N \ {\rm is \ odd}).
\end{align*}
\end{lemma}
\proof
When $N$ is even,
\[
\frac{4\pi}{N} {\mathbb Z} = \frac{2\pi}{N/2} {\mathbb Z} = 
\left\{ \frac{2\pi m}{N/2} \colon 0 \le m \le \frac{N}{2}-1 \right\} 
=\left\{ \frac{4\pi m}{N} \colon 0 \le m \le \frac{N}{2}-1 \right\}
\]
 in modulo $2\pi$.
 
When $N$ is odd, 
\[
\frac{4\pi}{N} \cdot \frac{N+1}{2} = \frac{2\pi}{N} \quad ({\rm mod}\ 2\pi).
\]
This implies 
\[
\frac{4\pi }{N} {\mathbb Z} = \frac{2\pi }{N} {\mathbb Z} = 
\left\{ \frac{2\pi m}{N} \colon 0 \le m \le N-1 \right\}
\]
in modulo $2 \pi$.
\endproof

\begin{lemma}\label{lemma3.3}
For any $\beta, \gamma, \delta \in {\mathbb R}$, there exist real numbers $\alpha$ and $l$ 
which satisfy
$N\alpha = \beta$ $({\rm mod} \ 2\pi)$, $Nl = \gamma$ $({\rm mod} \ 2\pi)$, 
$0 \le \delta + \alpha +2l < \frac{2\pi}{N}$ in modulo $2\pi$ and 
\[
0 \le \alpha < \frac{4\pi}{N}\quad ({\rm when}\ N \ {\rm is \ even}), \qquad 
0 \le \alpha < \frac{2\pi}{N} \quad ({\rm when}\ N \ {\rm is \ odd}).
\]
\end{lemma}

\proof
For the conditions $N\alpha = \beta$ and $Nl = \gamma$, 
$\alpha $ and $l$ should be
\[
\alpha = \frac{\beta}{N} + \frac{2\pi m_1}{N}, \qquad
l = \frac{\gamma}{N} + \frac{2\pi m_2}{N}
\]
for some $m_1, m_2 \in {\mathbb Z}$. Here,
\[
\delta + \alpha + 2l = \delta + \frac{\beta+2\gamma}{N} + \frac{2\pi m_1}{N} +\frac{4\pi m_2}{N}.
\]

When $N$ is odd, there exists $m_1 \in {\mathbb Z}$ such that
\[
0 \le \frac{\beta}{N} + \frac{2\pi m_1}{N} < \frac{2\pi}{N}.
\]
Moreover, by the previous lemma, there exists $m_2 \in {\mathbb Z}$ such that
\[
0\le  \delta + \frac{\beta+2\gamma}{N} + \frac{2\pi m_1}{N} +\frac{4\pi m_2}{N} < \frac{2\pi}{N}
\]
in modulo $2\pi$.

When $N$ is even, there exists $m_3 \in {\mathbb Z}$ such that
\[
0 \le \frac{\beta}{N} + \frac{2\pi m_3}{N} <
\frac{2\pi}{N} 
\le \frac{\beta}{N} + \frac{2\pi (m_3+1)}{N} < \frac{4\pi}{N}.
\]
Moreover, by the previous lemma, there exists $m_2 \in {\mathbb Z}$ such that
\[
-\frac{2\pi}{N} 
\le  \delta + \frac{\beta+2\gamma}{N} + \frac{2\pi m_3}{N} +\frac{4\pi m_2}{N} =:t
< \frac{2\pi}{N}
\]
in modulo $2\pi$. 
If $-\frac{2\pi}{N} \le t <0$, we set $m_1 = m_3+1$.
If $ 0\le t < \frac{2\pi}{N}$, we set $m_1 = m_3$.
Then, we obtain the assertion.
\endproof

\begin{lemma}\label{lem3.4}
Let $l$ be a real number which satisfies $lN=0$ $({\rm mod}\ 2\pi)$.
When $N$ is even,
\[
2l = \frac{4\pi m}{N}
\]
for some $m \in \{0,1, \ldots , N/2-1\}$ in modulo $2\pi$. When $N$ is odd,
\[
2l = \frac{2\pi m}{N}
\]
for some $m \in \{0, 1, \ldots N-1\}$
in modulo $2\pi$.
\end{lemma}
\proof
By the assumption, $l \in \frac{2\pi}{N} {\mathbb Z}$.
Hence, $2l \in \frac{4\pi}{N} {\mathbb Z}$.
Then, we have the assertion by Lemma \ref{lemma3.2}.
\endproof

In quantum mechanics, a state $\psi$ in a Hilbert space is identified with $e^{\im l} \psi$.
Moreover, almost all properties of $e^{\im l}U$, such as the spectrum and 
the distribution for an initial state, can be obtained from those of $U$.
Thus, we also identify a quantum walk $U$ with $e^{\im l}U$.

\begin{theorem}\label{thm3.5}
A quantum walk $U$ on a cycle is unitarily equivalent to
\begin{equation}\label{eq:thm3.4.2}
U_{r, \theta, \alpha} =\sum_{x\in V} 
\left( |\e_1^{x+1}\> \<r_x \e_1^x +s_x e^{\im \theta_x} \e_2^x |
+ |\e_2^{x-1}\> \<-s_x e^{\im(-\theta_x +\alpha)} \e_1^x +r_x e^{\im \alpha} \e_2^x |\right)
\end{equation}
for some $0 \le r_x \le 1$, $s_x = \sqrt{1-r_x^2}$, $\theta_1 =0$, $0\le \theta_2 < \frac{2\pi}{N}$,
$0 \le \theta_x <2\pi$ $(x = 3,4,\ldots N)$ and
\[
0 \le \alpha < \frac{4\pi}{N}\quad ({\rm when}\ N \ {\rm is \ even}), \qquad 
0 \le \alpha < \frac{2\pi}{N} \quad ({\rm when}\ N \ {\rm is \ odd}).
\]
\end{theorem}

\proof
We already show that $U$ is unitarily equivalent to
\[
U_\zeta 
= \sum_{x\in V} \left( |\e_1^{x+1}\> \<r_x e^{\im a_x} \e_1^x + s_x e^{\im b_x} \e_2^x| 
+ |\e_2^{x-1}\>\<  s_x e^{\im c_x} \e_1^x + r_x e^{\im d_x} \e_2^x | \right).
\]
Let $\alpha$ be a real number which satisfies
\begin{equation}\label{eq:3.13}
N\alpha = \sum_{k=1}^N d_k - \sum_{k=1}^N a_k \quad ({\rm mod}\ 2\pi),
\end{equation}
and let $l$ be a real number which satisfies
\begin{equation}\label{eq:3.14}
Nl = \sum_{k=1}^N a_k  \quad ({\rm mod}\ 2\pi).
\end{equation}
Define a unitary $W_x$ on $\H_x$ by
\[
W_x = \begin{bmatrix} e^{\im p_x} & 0 \\ 0 & e^{\im q_x} \end{bmatrix}
= e^{\im p_x} |\e_1^x\>\<\e_1^x| + e^{\im q_x}  |\e_2^x\>\< \e_2^x|,
\]
where
\begin{align*}
p_1&= 0, \quad p_x = \sum_{k=1}^{x-1} a_k -(x-1) l  \quad (2 \le x \le N), \\
q_1&= a_1-b_1, \quad q_x = -\sum_{k=2}^x d_k + (x-1)( \alpha + l) + a_1 - b_1 \quad (2 \le x \le N).
\end{align*}
Then, $W = \bigoplus_{x\in V} W_x$ is a unitary on $\H = \bigoplus \H_x$.
By simple calculation,
\begin{align}
e^{\im l} WU_\zeta W^* 
=& e^{\im l} \sum_{x\in V} \left( |W\e_1^{x+1}\> \<r_x e^{\im a_x} W \e_1^x + s_x e^{\im b_x}  W \e_2^x| 
+ |W \e_2^{x-1}\>\<  s_x e^{\im c_x} W \e_1^x + r_x e^{\im d_x} W\e_2^x | \right) \nonumber \\
=& 
\sum_{x\in V} \left( |\e_1^{x+1}\> \<r_x e^{\im (a_x+p_x -p_{x+1} -l)} \e_1^x 
+ s_x e^{\im (b_x +q_x -p_{x+1}-l)} \e_2^x| \right. \nonumber \\
& \left. + |\e_2^{x-1}\>\<  s_x e^{\im (c_x+p_x - q_{x-1} -l) } \e_1^x 
+ r_x e^{\im (d_x + q_x- q_{x-1} -l)} \e_2^x | \right). \label{eq:thm3.4}
\end{align}
By paying attention to the cases $x=1$ and $x=N$, we have
\begin{align*}
a_x + p_x -p_{x+1}-l &= 0 \quad (1 \le x \le N), \\
d_x + q_x -q_{x-1}-l &= \alpha  \quad (1\le x \le N).
\end{align*}
We set $\theta_x= b_x + q_x -p_{x+1} -l$. Then, $\theta_1 = 0$ and
\begin{align}
\theta_x = -\sum_{k=2}^x a_k -\sum_{k=2}^x d_k +b_x -b_1 + (x-1)(\alpha +2l) 
\quad (2\le x \le N). \label{eq3.5.1}
\end{align}
In particular, $\theta_2 = -a_2 -d_2 +b_2 -b_1 +\alpha +2l$.
By lemma \ref{lemma3.3}, there exist real numbers $\alpha$ and $l$ such that
$N\alpha = \sum_{k=1}^N d_k - \sum_{k=1}^N a_k$, 
$Nl = \sum_{k=1}^N a_k$,
\[
0 \le -a_2 -d_2 +b_2 -b_1 +\alpha +2l < \frac{2\pi}{N}
\]
in modulo $2\pi$ and 
\begin{align*}
0 \le \alpha < \frac{4\pi}{N}\quad ({\rm when}\ N \ {\rm is \ even}), \qquad
0 \le \alpha  < \frac{2\pi}{N} \quad ({\rm when}\ N \ {\rm is \ odd}).
\end{align*}
Since the vectors 
\[
r_x e^{\im (a_x+p_x -p_{x+1} -l)} \e_1^x 
+ s_x e^{\im (b_x +q_x -p_{x+1}-l)} \e_2^x \quad {\rm and} \quad
 s_x e^{\im (c_x+p_x - q_{x-1} -l) } \e_1^x 
+ r_x e^{\im (d_x + q_x- q_{x-1} -l)} \e_2^x 
\]
in \eqref{eq:thm3.4} make an orthonormal basis of $\H_x$,
\[
c_x + p_x -q_{x-1}-l = - \theta_x +\alpha +\pi.
\]
Consequently, 
\[
e^{\im l} WU_\zeta W^* 
= 
\sum_{x\in V} 
\left( |\e_1^{x+1}\> \<r_x \e_1^x +s_x e^{\im \theta_x} \e_2^x |
+ |\e_2^{x-1}\> \<-s_x e^{\im(-\theta_x +\alpha)} \e_1^x +r_x e^{\im \alpha} \e_2^x |\right)
=
U_{r, \theta,\alpha}.
\]
Therefore, we conclude that a quantum walk $U$ is unitary equivalent to 
$U_{r, \theta, \alpha}$ 
for some $0 \le r_x \le 1$, $\theta_1 =0$, $0 \le \theta_2 < \frac{2\pi}{N}$,
$0 \le \theta_x <2\pi$ $(x = 3,4,\ldots N)$ and
\[
0 \le \alpha < \frac{4\pi}{N}\quad ({\rm when}\ N \ {\rm is \ even}), \qquad 
0 \le \alpha < \frac{2\pi}{N} \quad ({\rm when}\ N \ {\rm is \ odd}). 
\]
\endproof

\begin{theorem}\label{thm3.6}
Quantum walks $U_{r, \theta, \alpha}$ and $U_{r', \theta', \alpha'}$
with $0<r_x, r'_x <1$,  $\theta_1 =\theta'_1 =0$, 
$0 \le \theta_2, \theta_2' < \frac{2\pi}{N}$,
$0 \le \theta_x, \theta'_x <2\pi$ $(x = 3,4,\ldots N)$ and
\[
0 \le \alpha, \alpha' < \frac{4\pi}{N}\quad ({\rm when}\ N \ {\rm is \ even}), \qquad 
0 \le \alpha, \alpha' < \frac{2\pi}{N} \quad ({\rm when}\ N \ {\rm is \ odd})
\]
are unitarily equivalent if and only if, for all $1\le x \le N$,
\begin{equation}
r_x = r'_x, \quad \theta_x = \theta'_x \quad {\rm and} \quad \alpha = \alpha'. \label{eq:thm3.5.1}
\end{equation}
\end{theorem}

\proof
If \eqref{eq:thm3.5.1} holds, then $U_{r, \theta, \alpha} = U_{r', \theta', \alpha'}$.
Therefore, $U_{r, \theta, \alpha}$ and $U_{r', \theta', \alpha'}$ are unitarily equivalent.

Conversely, we assume that $U_{r, \theta, \alpha}$ and $U_{r', \theta', \alpha'}$
are unitarily equivalent, that is, there exist a unitary $W= \bigoplus_{x\in V} W_x$ on $\H$
and a real number $l$ such that
\[
e^{\im l} W U_{r, \theta, \alpha} W^* = U_{r', \theta', \alpha'}.
\]
First, we consider the equation
\[
P_{x\pm 1} e^{\im l} W U_{r, \theta, \alpha} W^*P_x  =P_{x\pm 1} U_{r', \theta', \alpha'} P_x.
\]
By \eqref{eq:thm3.4.2},
\[
P_{x + 1} e^{\im l} W U_{r, \theta, \alpha} W^*P_x
= e^{\im l} |W \e_1^{x+1}\> \< W(r_x\e_1^x + s_x e^{\im \theta_x} \e_2^x) |
\]
and
\[
P_{x+1} U_{r', \theta', \alpha'} P_x
= |\e_1^{x+1}\> \< r'_x\e_1^x + s'_x e^{\im \theta'_x} \e_2^x |.
\]
Therefore, $\Ran (P_{x + 1} e^{\im l} W U_{r, \theta, \alpha} W^*P_x) = {\mathbb C} W \e_1^{x+1}$
and $\Ran ( P_{x+1} U_{r', \theta', \alpha'} P_x) ={\mathbb C} \e_1^{x+1}$,
so that $W\e_1^{x+1}  \in {\mathbb C} \e_1^{x+1}$.
Similarly, the equations
\begin{align*}
P_{x - 1} e^{\im l} W U_{r, \theta, \alpha} W^*P_x
&= e^{\im l} |W \e_2^{x-1}\> \< W(-s_xe^{\im(-\theta_x+\alpha)}\e_1^x 
+ r_x  e^{\im \alpha} \e_2^x) |\\
P_{x-1} U_{r', \theta', \alpha'} P_x
&= |\e_2^{x-1}\> \< -s'_x e^{\im(-\theta'_x +\alpha')} \e_1^x + r'_x e^{\im \alpha'} \e_2^x |
\end{align*}
imply $W\e_2^{x-1}  \in {\mathbb C} \e_2^{x-1}$.
Since $W=\bigoplus_{x\in V} W_x$ is unitary, $W_x$ can be written as
\[
W_x = \begin{bmatrix} e^{\im p_x} & 0 \\ 0 & e^{\im q_x} \end{bmatrix}
= e^{\im p_x} |\e_1^x\>\<\e_1^x| + e^{\im q_x}  |\e_2^x\>\< \e_2^x|
\]
for some $p_x, q_x \in {\mathbb R}$.

Now, we consider the equation $e^{\im l} W U_{r, \theta, \alpha} W^* = U_{r', \theta', \alpha'}$.
Because $(e^{\im t}W)  U_{r, \theta, \alpha} (e^{\im t}W)^* = W U_{r, \theta, \alpha} W^*$ 
for any $t \in {\mathbb R}$, we can assume that $p_1 =0$.
By simple calculation, we have
\begin{align*}
&e^{\im l} W U_{r, \theta, \alpha} W^* \\
&=
e^{\im l} \sum_{x\in V} 
\left( |W\e_1^{x+1}\> \<r_x W\e_1^x +s_x e^{\im \theta_x} W\e_2^x |
+ |W\e_2^{x-1}\> \<-s_x e^{\im(-\theta_x +\alpha)} W\e_1^x +r_x e^{\im \alpha} W\e_2^x |\right) \\
&= \sum_{x\in V} 
\left( |\e_1^{x+1}\> \<r_x e^{\im(p_x-p_{x+1}-l)}\e_1^x +s_x e^{\im (\theta_x+q_x -p_{x+1}-l)} \e_2^x |
\right. \\
& \quad \left. + |\e_2^{x-1}\> \<-s_x e^{\im(-\theta_x +\alpha+p_x-q_{x-1}-l)} \e_1^x +
r_x e^{\im (\alpha+q_x -q_{x-1} -l)} \e_2^x |\right).
\end{align*}
On the other hand,
\[
 U_{r', \theta', \alpha'}
 = \sum_{x\in V} 
\left( |\e_1^{x+1}\> \<r'_x \e_1^x +s'_x e^{\im \theta'_x} \e_2^x |
+ |\e_2^{x-1}\> \<-s'_x e^{\im(-\theta'_x +\alpha')} \e_1^x +r_x e^{\im \alpha'} \e_2^x |\right).
\]
Therefore, we get $r_x = r'_x$ and the equations
\begin{align*}
p_x-p_{x+1}-l =0, \quad & \theta_x+q_x -p_{x+1}-l =\theta'_x \\
-\theta_x +\alpha+p_x-q_{x-1}-l =-\theta'_x +\alpha' , \quad 
& \alpha+q_x -q_{x-1} -l = \alpha'
\end{align*}
in modulo $2 \pi$. By $p_1=0$ and the first equation, 
\[
p_x = -l(x-1) \quad (1\le x \le N).
\]
Moreover, $p_N -p_0 -l=0$ implies $lN =0$.
By the second equation with $x=1$, we have
$q_1 = 0$,
because $\theta_1 =\theta'_1 =0$. 
Then, by the fourth equation,
\[
q_x = (l-\alpha +\alpha')(x-1) \quad (1\le x \le N).
\]
Furthermore, $\alpha + q_1-q_N-l = \alpha'$ implies $(l-\alpha + \alpha') N = 0$.
The second equation is calculated as
\begin{equation}\label{eq:3.12}
\theta_x + (2l-\alpha + \alpha')(x-1) = \theta'_x \quad (1\le x \le N).
\end{equation}
In particular, 
\begin{equation}\label{eq:3.11}
\theta_2 + 2l - \alpha + \alpha' = \theta'_2.
\end{equation}
Since $lN = 0$ and $(l-\alpha+\alpha')N =0$,
\[
(\theta_2 - \theta'_2)N = 0.
\]
By the assumption $0\le \theta_2, \theta_2' < \frac{2\pi}{N}$, we obtain $\theta_2 = \theta_2'$.
Then, \eqref{eq:3.11} is
\[
 2l - \alpha + \alpha' =0.
\]
Here, $\alpha$ and $\alpha'$ satisfy
\[
-\frac{4\pi}{N} < \alpha - \alpha' < \frac{4\pi}{N}\quad ({\rm when}\ N \ {\rm is \ even}), \quad 
-\frac{2\pi}{N} < \alpha - \alpha' < \frac{2\pi}{N} \quad ({\rm when}\ N \ {\rm is \ odd}).
\]
Therefore, we obtain $\alpha-\alpha'= 2l =0$ by Lemma \ref{lem3.4}, and
hence, $\alpha = \alpha'$.
Moreover, by \eqref{eq:3.12}, 
$\theta_x = \theta'_x$ for all $1\le x \le N$.
Consequently, we conclude $\alpha = \alpha'$, $r_x= r'_x$ and $\theta_x=\theta'_x$ 
for all $1\le x \le N$.
\endproof

Theorem \ref{thm3.5} and \ref{thm3.6} say that
the unitary equivalence classes of quantum walks on a cycle are
parametrized by $\alpha, r_x$ and $\theta_x$.
As $N$ goes to infinity, the limits of $\alpha$ and $\theta_2$ are $0$.
On the other hand, a one-dimensional quantum walk is unitarily equivalent to
\[
\sum_{x\in{\mathbb Z}} \left( |\e_1^{x+1} \>\< r_x \e_1^x + e^{\im \theta_x} s_x \e_2^x|
+  |\e_2^{x-1} \>\< - e^{-\im \theta_x} s_x \e_1^x + r_x \e_2^x| \right)
\]
where $0 \le r_x \le 1$, $\theta_0=\theta_1=0$ and 
$0 \le \theta_x <2\pi$ $(x \neq 0,1)$.
Therefore, the parametrization of the unitary equivalence classes of one-dimensional quantum walks
is similar to that of quantum walks on a cycle with $N \to \infty$.

There is a natural shift operator $S$ on $\H$, that is,
\[
S \e_i^x = \e_i^{x+1}
\]
for $i=1,2$ and $x \in V$.
A quantum walk $U$ on a cycle is called translation-invariant if
\[
SUS^* = U.
\]
By the next theorem, unitary equivalence classes of translation-invariant quantum walk on
a cycle are parametrized by two real numbers.

\begin{corollary}
A translation-invariant quantum walk on a cycle is unitarily equivalent to
\[
U_{r, \alpha} = \sum_{x\in V}
\left( |\e_1^{x+1}\> \<r \e_1^x +s \e_2^x |
+ |\e_2^{x-1}\> \<-s e^{\im\alpha} \e_1^x +r e^{\im \alpha} \e_2^x |\right)
\]
for some $0 \le  r \le 1$, $s=\sqrt{1-r^2}$ and
\[
0 \le \alpha < \frac{4\pi}{N}\quad ({\rm when}\ N \ {\rm is \ even}), \qquad 
0 \le \alpha < \frac{2\pi}{N} \quad ({\rm when}\ N \ {\rm is \ odd}).
\]
Moreover, $U_{r,\alpha}$ and $U_{r', \alpha'}$ with $r, r' \neq 0,1$ are unitarily equivalent
if and only if $r=r'$ and $\alpha=\alpha'$.
\end{corollary}

\proof
We need to consider Theorem \ref{thm2.3} and \ref{thm2.6}, again.
By Theorem \ref{thm2.3}, $U$ can be written as
\[
U = \sum_{x\in V} \sum_{i=1}^2  |\xi_i^x\>\<\zeta_i^x|.
\]
Since $U$ is translation-invariant, there exist $\xi_i$ and $\zeta_i$ $(i=1,2)$ in ${\mathbb C}^2$ 
such that
$\xi_i^x = \xi_i$ and $\zeta_i^x = \zeta_i$ for all $x \in V$.
Hence, the unitary $W$ in \eqref{eq2.6} is described as
\[
W = \bigoplus_{x\in V} \sum_{i=1}^2 |\e_i^x\> \<\xi_i^x|
\]
and is translation-invariant.
Therefore, $U_\zeta = WUW^*$ in Theorem \ref{thm2.6} is also translation-invariant and is written as 
\[
U_\zeta = \sum_{x\in V} \left( |\e_1^{x+1}\> \<\eta_1^x| + |\e_2^{x-1}\>\< \eta_2^x | \right),
\]
where $\eta_1^x = \eta_1$ and $\eta_2^x = \eta_2$
for some $\eta_1, \eta_2 \in {\mathbb C}^2$, as we see in \eqref{eq3.1}.
This implies that there exist $0\le r \le 1$ and $a,b,c,d \in {\mathbb R}$ such that,
for all $x\in V$, $r_x=r$, $a_x=a$ and so on.
Then, $\theta_x$ in \eqref{eq3.5.1} is
\[
\theta_x = (x-1)(-a-d+\alpha +2l) = (x-1) \theta_2.
\]
By \eqref{eq:3.13} and \eqref{eq:3.14}, $N \theta_2 = 0$ $({\rm mod}\ 2\pi)$.
Since $0\le \theta_2 < \frac{2\pi}{N}$, $\theta_2 =0$, and hence, $\theta_x =0$ $(1\le x \le N)$.
Consequently, $U_\zeta$ is unitarily equivalent to $U_{r, \alpha}$.

The remaining assertion follows from Theorem \ref{thm3.6}, immediately.
\endproof

\bigskip
\noindent
{\bf Acknowledgement.} 
This work was supported by 
JSPS KAKENHI Grant Numbers 17K05274.


\end{document}